\documentclass{ifacconf}

\usepackage{graphicx}      
\usepackage{natbib}        

\newif\ifIFAC
\IFACtrue

\usepackage{graphicx}

\usepackage[cmex10]{amsmath}
\usepackage{amsfonts}
\usepackage{amssymb}
\usepackage{dsfont}
\usepackage{accents}
\usepackage{url}

\usepackage{empheq}
\usepackage[normalem]{ulem}

\usepackage{enumerate}

\usepackage{tikz} 
\usepackage{pgfplots}
\usetikzlibrary{patterns}

\DeclareMathOperator*{\argmin}{arg\,min}
\DeclareMathOperator*{\argmax}{arg\,max}
\DeclareMathOperator{\st}{s.t.}

\newcommand{\ind}[1]{\text{I}\{#1\}}

\newcommand{\E}[1]{\mathbb{E}\big\{#1\big\}}

\usepackage{mathtools}
\DeclarePairedDelimiter{\PrfencesHard}{[}{]}

\DeclarePairedDelimiter{\PrfencesSoft}{(}{)}
\renewcommand{\Pr}{\operatorname{Pr}\PrfencesHard}

\renewcommand{\vec}{\operatorname{vec}\PrfencesSoft}

\newcommand{\fref}[1]{Fig.~\ref{#1}}
\newcommand{\tref}[1]{Table~\ref{#1}}
\newcommand{\eref}[1]{(\ref{#1})}
\newcommand{\sref}[1]{Section~\ref{#1}}

\newcommand{\aref}[1]{Appendix~\ref{#1}}

\usepackage{xcolor}
\newcommand{\comment}[1]{}  

\ifIFAC
\usepackage{enumitem} 
\else

\makeatletter%
\@ifclassloaded{IEEEconf}{
\makeatother%

}{}

\usepackage{amsthm}
\fi

\newtheorem{remark}{Remark}

\newcommand{\Eb}{{\mathbb E}}
\newcommand{\Rb}{{\mathbb R}}

\usepackage{dsfont}

\newcommand{\tikzmark}[1]{\tikz[overlay,remember picture] \node (#1) {};}
\tikzset{square arrow/.style={to path={-- ++(0,-.27) -| (\tikztotarget)}}}

\usepackage{subcaption}

\usepackage{empheq}
\usepackage{subcaption}
\usepackage{bm} 

\usepackage{balance}

\pgfplotsset{select coords between index/.style 2 args={
    x filter/.code={
        \ifnum\coordindex<#1\fi
        \ifnum\coordindex>#2\fi
    }
}}

\allowdisplaybreaks[1]

\begin{document}
\begin{frontmatter}

\title{Computing monotone policies for Markov decision processes: a nearly-isotonic
penalty approach\thanksref{footnoteinfo}} 


\thanks[footnoteinfo]{This work has been accepted for presentation at the 20th World
    Congress of the International Federation of Automatic Control, 9-14 July 2017. This
    work was partially supported by the Swedish Research Council under contract 2016-06079
    and the Linnaeus Center ACCESS at KTH.}

\author[KTH]{Robert Mattila} 
\author[KTH]{Cristian R. Rojas} 
\author[Cornell]{Vikram Krishnamurthy}
\author[KTH]{Bo Wahlberg}

\address[KTH]{Department of Automatic Control,
School of Electrical Engineering, KTH Royal Institute of Technology.
Stockholm, Sweden. \\ (e-mails: rmattila@kth.se, crro@kth.se, bo.wahlberg@ee.kth.se).}
\address[Cornell]{Department of
    Electrical and Computer Engineering, Cornell University. Cornell Tech, NY, USA.
    (e-mail: vikramk@cornell.edu)}

\begin{abstract}                
    This paper discusses algorithms for solving \emph{Markov decision
    processes} (MDPs) that have monotone optimal policies. We propose
    a two-stage alternating convex optimization scheme that can accelerate the
    search for an optimal policy by exploiting the monotone property. The
    first stage is a linear program formulated in terms of the joint
    state-action probabilities. The second stage is a regularized problem
    formulated in terms of the conditional probabilities of actions given
    states. The regularization uses techniques from \emph{nearly-isotonic
    regression}.  While a variety of iterative method can be used in the first
    formulation of the problem, we show in numerical simulations that, in particular, the
    \emph{alternating method of multipliers} (ADMM) can be significantly
    accelerated using the regularization step.
\end{abstract}


\end{frontmatter}

\section{Introduction}

Supermodularity conditions that ensure that a \emph{Markov decision process}
(MDP) has a monotone optimal policy have been studied widely (see, e.g.,
\cite{puterman_markov_1994} or \cite{krishnamurthy_partially_2016}, and
references therein).  In particular, such monotone policies provide a sparse
characterization of the optimal policy when the action space is small and the
state-space is large.  Computing the optimal policy for such an MDP is
computationally demanding; for example, the value iteration algorithm involves
$\mathcal{O}(X^2 U)$ computations per iteration, where $X$ and $U$ are the
number of states and actions, respectively. We aim to exploit structural
properties of this type of MDP to reduce the computational burden and hence
accelerate the search for an optimal policy. 

It was noted in \cite{krishnamurthy_computing_2013} that a monotone policy has
a piecewise constant structure. In the case of infinite-horizon MDPs, it was
shown how the search for a stationary optimal policy could be significantly
accelerated by means of techniques from sparse estimation (in particular, LASSO
techniques -- see, e.g., \cite{hastie_elements_2013}). In this paper, we build
upon this work but with two important differences: firstly, we generalize to
finite-horizon MDPs, for which it is even more important to exploit sparsity to
reduce the computational cost since they have non-stationary policies, and
secondly, we exploit the monotonicity explicitly, instead of only implicitly
through the piecewise constant property of monotone policies. 

The search for an optimal policy can be formulated as a \emph{linear program}
(LP). This LP can be solved in various iterative ways. The key idea in this
paper is to accelerate the iterative search by using the recently proposed
\emph{nearly-isotonic regression} technique by
\cite{tibshirani_nearly-isotonic_2011}. Roughly, a penalty is attached to
non-monotone iterates by means of adding an $l_1$ rectifier-like regularizer to
the cost function:
\begin{equation}
    \max\{0, f(x) - f(x+1)\},
\end{equation}
where $x$ is the iterate and $f$ is a monotone function of $x$.

This promotes monotonicity in the iterates because, intuitively, the regularization term
modifies the cost surface to be more steep in the direction of monotone policies -- where
the optimum is located. Unfortunately, including this regularizer in the LP is, first of
all, not straight-forward and secondly, it yields a non-linear, possibly non-convex,
problem.  We perform a relaxation of the regularized problem to obtain a convex problem.
Our method allows us to swap between the original LP formulation and the regularized
formulation, and thus take advantage of both by alternating between which one is used to
update the iterate: the original LP formulation guarantees convergence to the global
optimum and the regularized convex formulation accelerates the search by exploiting
monotonicity.

The main contributions of this paper are three-fold:
\begin{itemize}
    \item We show in numerical simulations that the number of iterations needed
        to converge can be vastly reduced using the regularization. This can
        yield a significant speed-up for large-scale systems where each
        iteration is time-consuming to calculate.
    \item The benefits are shown to be larger when a tuning parameter in the
        algorithm used to iteratively solve the LP is not chosen optimally.
        Since the optimal choice of this parameter is not known \emph{a
        priori}, it provides robustness to the search algorithm.
    \item Even though we generalize to finite-horizon MDPs, the method is
        directly applicable to infinite-horizon MDPs, and the
        regularizer is stronger (monotone) than the one in
        \cite{krishnamurthy_computing_2013} (piecewise constant). It can as
        such be seen as a direct improvement.
\end{itemize}

The outline of the paper is as follows.  We present preliminaries related to
MDPs in \sref{sec:prelim}, and then proceed with a discussion of the objective
as well as related work in \sref{sec:problem_formulation}.
\sref{sec:algorithm} presents the algorithm. Conditions for an MDP to have
monotone structure, as well as real-world examples, are presented along with
numerical simulations in \sref{sec:examples}. The paper is then concluded with
a brief summary and indications for future work in \sref{sec:conclusions}.

\section{Preliminaries}
\label{sec:prelim}

We let $\ind{\cdot}$ denote the indicator function. For a matrix $A$, define
$A(i,:)$ to be the $i$th row and $A(:,j)$ to be the $j$th column. We use the
corresponding slicing notation for higher order arrays. Let $\{x\}_+ = \max\{0,
x\}$ denote the positive part of a number $x$. In this paper, we use the words
\emph{monotone}, \emph{decreasing} and \emph{increasing} in the weak sense,
e.g., increasing means non-decreasing. The $l_\infty$-norm of a vector $v$ is
$\|v\|_\infty = \max_k |v_k|$, and $\|v\|_2$ denotes the standard Euclidean
norm. 

\subsection{Markov decision processes}
\label{sec:mdps}

Let $k = 0, 1, \dots, N$ denote discrete time. A \emph{Markov decision
process} (MDP) is a controlled Markov chain with state-space $\mathcal{X}
= \{1, 2, \dots, X\}$ and state $x_k \in \mathcal{X}$ at time $k$. It is
controlled in the sense that the transition matrices
\begin{equation}
    P_{ij}(u,k) = \Pr{x_{k+1} = j | x_k = i, u_k = u},
\end{equation}
are functions of time $k$ and action $u \in \mathcal{U} = \{1, 2, \dots, U\}$.
Associated with every state $i$, action $u$, and time $k$ is an immediate cost $c(i,u,k)$.
We consider a time horizon of length $N$ and assume that the terminal cost
$c(i,u,N) = c_N(i)$ is independent of action.  The aim of
the MDP is to find a policy $\bm \mu = \{\mu_0, \mu_1, \dots, \mu_{N-1}\}$,
where each $\mu_k$ is a mapping from the state-space to a (possibly degenerate)
probability distribution over the action set. In particular, the sought
policy is an \emph{optimal} policy, i.e., one such that
\begin{equation}
    \bm \mu^* = \argmin_{\bm \mu} J_{\bm \mu}(x),
    \label{eq:def_optimal_policy}
\end{equation}
for all initial states $x$, where 
\begin{equation}
    J_{\bm \mu}(x) = \Eb \Bigg\{\sum_{k=0}^{N-1} c(x_k, u_k, k) 
    + c_N(x_N) \big| x_0 = x \Bigg\}
\end{equation}
is the finite-horizon objective (expected cumulative cost incurred by $\bm
\mu$), and $u_k$ is distributed according to $\mu_k(x_k)$.

A policy is said to be deterministic at time $k$ if the probability
distribution induced by $\mu_k$ on the action space for each state $x$ is
degenerate, i.e., the probability mass in concentrated on one action. A policy
is said to be \emph{monotone} at time $k$ if the function
\begin{equation}
    \tilde \mu_k(x) = \E{u_k | x_k = x}
    \label{eq:tilde_mu}
\end{equation}
is monotone in $x$.\footnote{In this paper, we consider only monotonically
increasing policies, and will use the words \emph{monotone} and
\emph{increasing} interchangeably.} Note that this reduces to the standard
definition when considering deterministic policies. An MDP is said to have
a \emph{monotone optimal policy} if there is an optimal policy that is monotone
for each time $k$.

Motivated by problems in telecommunications and safety critical planning, see,
e.g., \cite{altman_constrained_1999}, \cite{krishnamurthy_partially_2016} and
\cite{el_chamie_convex_2016}, we allow for average-type constraints in the
problem:
\begin{equation}
    \mathbb{E}\Big\{ \sum_{k=0}^N \beta_l(x_k, u_k, k) \Big\} \leq
    \gamma_l \quad \text{ for } l = 1, \dots, L,
    \label{eq:mdp_constraints}
\end{equation}
where the $L$ functions $\beta_l(x,u,k)$ and thresholds $\gamma_l$ are given.
We refer to solving \eref{eq:def_optimal_policy}, subject to the constraints
\eref{eq:mdp_constraints}, as the \emph{constrained case} when $L > 0$.

The search for an optimal policy, i.e., problem \eref{eq:def_optimal_policy} (with or
without the constraints \eref{eq:mdp_constraints}), can be approached in different ways --
see, e.g., \cite{puterman_markov_1994} or \cite{krishnamurthy_partially_2016}. One
possibility is to formulate the optimality conditions as a \emph{linear program} (LP).
This has the benefit of facilitating sensitivity analysis of the obtained solution, and
also facilitating the inclusion of constraints, such as \eref{eq:mdp_constraints}, in the
problem.

Assume $x_0$ to be the initial state of the MDP. Then an optimal policy can be
found using the following LP\footnote{Although in classical textbooks,
    infinite-horizon MDPs are solved via linear programming, it is
    straightforward to formulate the solution of a finite-horizon MDP as an LP.} (see, e.g., \cite[Chapter 12]{feinberg_handbook_2002} for details):
\begin{align}
    \allowdisplaybreaks
    \min_{\substack{\pi \in \\ \Rb^{X U (N+1)}}} & \sum_{x \in \mathcal{X}} \sum_{u
\in \mathcal{U}} \Big\{ \sum_{k=0}^{N-1} c(x,u,k) \pi(x,u,k) \notag \\
& \hspace{1.85cm} + c_N(x) \pi(x,u,N) \Big\} \notag \\
    \text{s.t.} \hspace{0.5cm}  & \sum_{u \in \mathcal{U}} \pi(x,u,0) = \ind{x = x_0}
    \quad \text{ for } x \in \mathcal{X}, \notag \\
    & \sum_{u \in \mathcal{U}} \pi(j,u,k) = \sum_{i \in \mathcal{X}}
    \sum_{u \in \mathcal{U}} P_{ij}(u,k) \pi(i,u,k-1) \notag \\
    & \hspace{3.28cm} \text{ for } j \in \mathcal{X}, k = 1,2,\dots,N, \notag \\
    & \pi(x,u,k) \geq 0 \;\; \text{ for } x \in \mathcal{X}, u \in \mathcal{U},
    k = 0,1,\dots,N, \notag \\
    & \sum_{x \in \mathcal{X}} \sum_{u \in \mathcal{U}} \sum_{k=0}^N \pi(x,u,k)
    \beta_l(x,u,k) \leq \gamma_l \notag \\ 
    & \hspace{3.5cm} \text{ for } l = 1,2, \dots, L.
    \label{eq:LP}
\end{align}
In this formulation, $\pi$ is an occupation measure, namely: 
\begin{equation}
    \pi(x,u,k) = \Pr{x_k = x, u_k = u}.
\end{equation}
The associated policy $\bm \mu^*$ is 
\begin{equation}
    u_k^*(x) = u \text{ with probability } \theta(x,u,k),
\end{equation}
where the conditional probabilities $\theta(x,u,k) = \Pr{u_k = u | x_k = x}$ can be
calculated as
\begin{equation}
    \theta(x,u,k) = \frac{\pi(x,u,k)}{\sum_{\bar u \in \mathcal{U}} \pi(x,\bar
    u,k)}.
    \label{eq:theta_def}
\end{equation}
\begin{remark}
It should be noted that in terms of these variables, we can re-write the
function defining monotonicity, i.e., equation \eref{eq:tilde_mu}, as
\begin{align}
    \tilde \mu_k(x) &= \E{u_k | x_k = x} \notag \\
                    &= \sum_{u=1}^U u \, \theta(x,u,k) \notag \\
                    &= [1\;2\;\cdots\;U] \; \theta(x,:,k).
    \label{eq:mu_tilde}
\end{align}
\end{remark}

\section{Problem Formulation \hspace{3cm} and Related Work}
\label{sec:problem_formulation}

If it is \emph{a priori} known that an MDP has an optimal policy $\bm \mu^*$
that is monotone -- see \sref{sec:examples} for examples where this holds --
then the question we aim to answer in this paper is: \emph{how can we
efficiently exploit the structure to find $\bm \mu^*$?}

The problem is of most interest when considering large-scale MDPs.  We first
note that a direct search for an optimal policy over the space of monotone
policies (which is vastly smaller than the complete policy space) is, in the
case of infinite-horizon MDPs, a combinatorial search over ${X+U-1 \choose
U-1}$ stationary policies. In the finite-horizon case, this increases to
${X+U-1 \choose U-1}^N$ non-stationary policies. This quickly becomes
prohibitively large.

Work on large-scale MDPs that does not explicitly take monotonicity into
account include \emph{approximate dynamic programming} (ADP), where, e.g., the
optimal value function $J_{\bm \mu^*}$ is approximated by a linear expansion in
some terms of some basis functions, and the related \emph{neuro-dynamic
programming}. See, e.g., \cite{bertsekas_dynamic_2007},
\cite{de_farias_linear_2003}, and \cite{bertsekas_neuro-dynamic_1995}.

In the recent work by \cite{fu_optimal_2015}, block-splitting methods (that are
based on the alternating direction method of multipliers, ADMM, which we also
use in this paper) are employed to solve large-scale MDPs by means of
decomposing the problem into sub-problems that can be solved in a distributed
fashion. We believe that their work could propitiously be used in conjunction
with the work presented in this paper (for monotone MDPs).

\cite{jiang_approximate_2015} provide an extensive review of real-world
applications of MDPs with monotone value functions, along with a method based
on ADP that exploits the monotonocity of the value function.
Our method in comparison promotes the monotonicity directly in the policy
space. 

In \cite{ngo_monotonicity_2010}, and see also
\cite{krishnamurthy_partially_2016}, it was proposed that the problem of
finding an optimal monotone policy can be relaxed by approximating the optimal
policy by a continuous representation based on sigmoidal functions. The search
for an optimal policy can then be approached using simulation based stochastic
optimization.

The most closely related work is \cite{krishnamurthy_computing_2013}. There, it
was proposed how the monotonicity of an optimal policy, in the infinite-horizon
case, can be exploited.  Since the action set is finite, the number of jumps
that the policy can make (as a function of state) is limited to at most $U-1$.
This implies that the policy is sparse in the number of jumps. It is natural to
exploit this structure by using methods from sparse estimation. In particular,
the \emph{(fused) group LASSO} by \cite{yuan_model_2006} was employed.  This,
however, promotes only a piecewise constant structure in the policy -- not
necessarily monotonicity, which is explicitly promoted in this work. Also, the
finite-horizon setup that is considered here results in a much larger problem
since the policy is non-stationary and hence it has more decision variables (a
factor $N$) in the corresponding LP.

\section{Isotonic Regularization \hspace{3cm} for Monotone MDP\footnotesize{s}}
\label{sec:algorithm}

As mentioned above, the key point is that for MDPs with large state space and small
action space, a monotone policy is sparse. We use an iterative optimization
algorithm to solve problem \eref{eq:LP} by exploiting sparsity. Assuming that
we know that there exists an optimal policy that is monotone (see
\sref{sec:examples} for conditions and examples), we employ the idea from
\cite{tibshirani_nearly-isotonic_2011}, but in a regularization setting.  

The key idea is to add a rectified $l_1$-penalty of the form 
\begin{equation}
    \sum_{x=1}^{X-1} \big\{ \tilde \mu_k (x) - \tilde \mu_k(x+1)\big\}_+
    \label{eq:regularization_term}
\end{equation}
to the cost in the optimization problem -- since the function $\tilde \mu_k(x)$
is assumed to be (monotonically) increasing in $x$ at the optimum. Intuitively,
this will modify the cost-surface to be more steep in the direction of monotone
policies -- resulting in faster convergence of the iterative optimization
algorithm.

However, there are difficulties performing this regularization. The main
difficulty is that it is not possible to directly add the term
\eref{eq:regularization_term} in the original LP \eref{eq:LP}. Adding it
involves a change of variables (\eref{eq:theta_def}-\eref{eq:mu_tilde}) that
turns the problem into a non-linear, and possibly non-convex, problem. Our
approach is an alternating optimization scheme, where we switch between
updating the iterate on the globally convergent LP formulation and the
regularized problem, and hence, exploit the benefits of both formulations. We
will show below the details of these two steps and how it is possible to
alternate between the two formulations.

\subsection{Linear program update}
\label{sec:ADMM}

There are several ways to iteratively solve an LP such as \eref{eq:LP}, see,
e.g., \cite{luenberger_linear_2008}. The accelerating regularization technique
that we demonstrate in this paper is applicable to any iterative method where
the iterates are not restricted to the vertices of the feasible domain. The
goal of the regularization is to decrease the number of iterations
needed until the iterates converge to an optimal monotone solution.

This motivates our choice to use the \emph{alternating direction method of
multipliers} (ADMM), see \cite{boyd_distributed_2010}, which is a popular
method to solve large-scale optimization problems.  Second-order methods (such
as interior-point methods) often converge in few iterations to very high
accuracy. However, for very large problems, even a single iteration of an
interior point method might be computationally infeasible. In comparison,
first-order methods and ADMM converge using a higher number of cheap
iterations.

The ADMM update equations for LPs have been derived in
\cite{boyd_distributed_2010}. To utilize these, we first put the problem on
standard LP form. It is straight-forward to re-write problem \eref{eq:LP}
using matrix-vector notation as 
\begin{align}
    \min_\alpha & \quad q^T \alpha \notag \\
    \st & \quad A\alpha = b, \notag \\
        & \quad \alpha \geq 0,
\end{align}
where $\alpha$ is a vectorized version of the decision variable $\pi$, and $q$, $A$
and $b$ follow from the cost and the constraints. In terms of these variables,
an ADMM update (from iteration $n$ to $n+1$) is obtained by first solving the set of linear equations
\begin{equation}
    \begin{bmatrix}
        \rho I & A^T \\
        A & 0
    \end{bmatrix}
    \begin{bmatrix}
        \alpha^{(n+1)} \\
        \nu
    \end{bmatrix}
    +
    \begin{bmatrix}
        q - \rho(z^{(n)} - \eta^{(n)}) \\
        -b
    \end{bmatrix}
    = 0,
    \label{eq:ADMM_iter_1}
\end{equation}
where $\rho > 0$ is the tuning parameter of ADMM, and then updating the dual
variables as
\begin{align}
    z^{(n+1)} &= \{\alpha^{(n+1)} + \eta^{(n)}\}_+ \; , \label{eq:ADMM_iter_2} \\
    \eta^{(n+1)} &= \eta^{(n)} + \alpha^{(n+1)} - z^{(n+1)}. \label{eq:ADMM_iter_3}
\end{align}

\begin{remark}
Another reason for using ADMM is apparent here: it has only one tuning
parameter, namely $\rho$. ADMM is moreover very generous in terms of
convergence with respect to this parameter, in fact, under mild conditions, it
is convergent for any choice of $\rho$, albeit the performance may vary --
this is explored in the numerical examples in \sref{sec:examples}.
\end{remark}

In terms of the ADMM variables, the \emph{primal residual} of the LP is
\begin{equation}
    r^{(n)} = \alpha^{(n)} - z^{(n)}.
\end{equation}
This is a measure of how feasible the current iterate is.

\subsection{Isotonic regularization}
\label{sec:isotonic_LP}

Recall from \sref{sec:mdps} that when an MDP has an optimal monotone policy,
the scalar function from equation \eref{eq:tilde_mu},
\begin{align}
    \tilde \mu_k(x) &= \E{u_k | x_k = x} \notag \\
                    &= \sum_{u=1}^U u \, \theta(x,u,k) \notag \\
                    &= [1\;2\;\cdots\;U] \; \theta(x,:,k),
\end{align}
is monotonically (increasing) in state $x$ for each time $k$ (if $\theta$
corresponds to an optimal policy).

A natural choice of regularization to include in the problem is thus the
following penalty introduced in a general regression setting in
\cite{tibshirani_nearly-isotonic_2011}:
\begin{align}
    \lambda &\sum_{k=0}^N \sum_{x=1}^{X-1} \Big\{ \tilde \mu_k(x) - \tilde
    \mu_k(x+1) \Big\}_+ = \notag \\
    \lambda &\sum_{k=0}^N \sum_{x=1}^{X-1} \Big\{ [1\;2\;\cdots\;U] \big( \theta(x,:,k)
    - \theta(x+1,:,k) \big) \Big\}_+,
    \label{eq:penalty}
\end{align}
where $\lambda$ is the regularization weight. This adds a penalty whenever the
iterate, i.e., the policy, is not monotone.

The main problem is that this regularization term is naturally formulated in
terms of the conditional probabilities $\theta(x,u,k)$, rather than the joint
occupation probabilities $\pi(x,u,k)$, in which the LP \eref{eq:LP} is
formulated. To deal with this, we reformulate problem \eref{eq:LP} as an
equivalent problem using the marginalized state probabilities and the
conditionals $\theta(x,u,k)$. The key is to derive a way to swap between these
two formulations.

To do this, introduce
\begin{equation}
    p(x,k) = \Pr{x_k = x}
\end{equation}
as the state distribution at time $k$. The relations we need to be able to
change formulation are equation \eref{eq:theta_def}, and the following two
relations;
\begin{align}
    p(x,k) &= \Pr{x_k = x} \notag \\
           &= \sum_{\bar u \in \mathcal{U}} \Pr{x_k = x, u_k = \bar u} \notag \\
            &= \sum_{\bar u \in \mathcal{U}} \pi(x, \bar u, k),
    \label{eq:pi2p}
\end{align}
    and
\begin{align}
    \pi(x,u,k) &= \Pr{x_k = x, u_k = u} \notag \\
               &= \Pr{u_k = u | x_k = x} \Pr{x_k = x} \notag \\
               &= \theta(x,u,k) p(x,k).
    \label{eq:pi2thetap}
\end{align}

In terms of $p$ and $\theta$, problem \eref{eq:LP} with an included
regularization term \eref{eq:penalty} reads
\begin{align}
    \min_{\substack{p \in \Rb^{X (N+1)} \\ \theta \in \Rb^{X
    U (N+1)}}} & \quad \sum_{x \in \mathcal{X}} \sum_{u
\in \mathcal{U}} \Big\{ \sum_{k=0}^{N-1} c(x,u,k) \theta(x,u,k) p(x,k) \notag \\
    & \hspace{1cm} + c_N(x) \theta(x,u,N) p(x,N) \Big\} \notag \\
    & \hspace{-0.9cm} + \lambda \sum_{k=0}^N \sum_{x=1}^{X-1} \Big\{ [1\;2\;\cdots\;U] \big( \theta(x,:,k)
    - \theta(x+1,:,k) \big) \Big\}_+ \notag \\
    \text{s.t.} \quad & \sum_{u \in \mathcal{U}} \theta(x,u,0) p(x,0) = \ind{x = x_0}
    \quad \text{ for } x \in \mathcal{X}, \notag \\
    & \sum_{u \in \mathcal{U}} \theta(j,u,k) p(j,k) = \notag \\
    & \hspace{1cm} \sum_{i \in \mathcal{X}}
    \sum_{u \in \mathcal{U}} P_{ij}(u,k) \theta(i,u,k-1) p(i,k-1) \notag \\
    & \hspace{3cm} \text{ for } j \in \mathcal{X}, k = 1,\dots,N, \notag \\
    & \sum_{x \in \mathcal{X}} \sum_{u \in \mathcal{U}} \sum_{k=0}^N
    \theta(x,u,k) p(x,k) \beta_l(x,u,k) \leq \gamma_l \notag \\
    & \hspace{4cm} \text{ for } l = 1,2, \dots, L, \notag \\
    & \theta(x,u,k) \geq 0 \; \text{ for } x \in \mathcal{X}, u \in
    \mathcal{U}, k = 0,1,\dots,N, \notag \\
    & \sum_{u \in \mathcal{U}} \theta(x,u,k) = 1 \quad \text{ for } x \in
    \mathcal{X}, k = 0,1,\dots,N, \notag \\
    & p(x,k) \geq 0 \quad \text{ for } x \in \mathcal{X}, k = 0,1, \dots, N,
    \notag \\
    & \sum_{x \in \mathcal{X}} p(x,k) = 1 \quad \text{ for } k = 0,1, \dots, N.
    \label{eq:full_regularized}
\end{align}

In order to simplify the regularization update step, we \emph{i)} assume $p$ to be fixed,
\emph{ii)} drop the redundant constraints, and \emph{iii)} relax the constraints related
to the initial distribution, the state transitions and the average-type constraints
\eref{eq:mdp_constraints}. This allows for more flexibility in the regularized update --
\emph{note that these will anyway be enforced later in the original LP formulation}. This
yields the relaxed problem
\begin{align}
    \min_{\substack{\theta \in \\ \Rb^{X U (N+1)}}} & \sum_{x \in \mathcal{X}} \sum_{u
\in \mathcal{U}} \Big\{ \sum_{k=0}^{N-1} c(x,u,k) \theta(x,u,k) p(x,k) \notag \\
    & \hspace{1.8cm} + c_N(x) \theta(x,u,N) p(x,N) \Big\} \notag \\
    &\hspace{-0.85cm} + \lambda \sum_{k=0}^N \sum_{x=1}^{X-1} \Big\{ [1\;2\;\cdots\;U] \big( \theta(x,:,k)
    - \theta(x+1,:,k) \big) \Big\}_+ \notag \\
    \text{s.t.} \hspace{0.5cm} 
    & \hspace{-0.2cm} \theta(x,u,k) \geq 0 \quad \text{ for } x \in \mathcal{X}, u \in
    \mathcal{U}, k = 0,1,\dots,N, \notag \\
    &\hspace{-0.2cm} \sum_{u \in \mathcal{U}} \theta(x,u,k) = 1 \quad \text{ for } x \in
    \mathcal{X}, k = 0,1,\dots,N.
    \label{eq:penalized_problem}
\end{align}

\subsection{Regularized subgradient step}

Again, the idea is that the regularization will promote monotonicity by
increasing the slope in the direction of monotone policies (where the optimum 
is located). However, the simplifications done to arrive at problem
\eref{eq:penalized_problem} probably shift the minimum of the optimization
problem away from the original minimum in problem \eref{eq:LP}. For this reason, we
need to return to the original (globally convergent) formulation and have the
effect of the regularization step diminish over time.

Therefore, and due to the non-smooth objective function, we employ 
the subgradient method, see \cite{nesterov_introductory_2004}.  The nominal
problem for the subgradient method is
\begin{align}
    \min_\beta & \quad f(\beta) \notag \\
    \st & \quad \beta \in Q, \notag \\
        & \quad \bar f(\beta) \leq 0,
    \label{eq:subgradient_nominal}
\end{align}
where $f$ is a cost function, $Q$ is a convex set and $\bar f$ is an inequality
constraint function. In our case, compare with problem
\eref{eq:penalized_problem}, we have that the decision variables $\beta$ are
the conditionals $\theta$, $f$ is the regularized cost function, $Q$ are
simplices for slices of $\beta$, and $\bar f$ is a negative equality mapping.

Denote a subgradient of the cost function $f$ as $g$ and a subgradient of the
inequality constraint function $\bar f$ as $\bar g$. The subgradient method
consists of the following two steps. At iteration $n$,
\begin{enumerate}
    \item Compute $f(\beta^{(n)})$, $g(\beta^{(n)})$, $\bar f(\beta^{(n)})$ and $\bar
        g(\beta^{(n)})$ and set
        \begin{equation}
            p^{(n)} = \begin{cases}
                g(\beta^{(n)}) \text{ if } \bar f(\beta^{(n)}) < \|\bar g(\beta^{(n)})
                    \|_2 \, \frac{R}{\sqrt{n + 0.5}}, \\
                    \bar g(\beta^{(n)}) \text{ if } \bar f(\beta^{(n)}) \geq \|\bar
                    g(\beta^{(n)}) \|_2 \, \frac{R}{\sqrt{n + 0.5}}, \\
                \end{cases}
                \label{eq:SG_iter_1}
        \end{equation}
    \item Set \begin{equation}
            \beta^{(n+1)} = \pi_Q\Big\{\beta^{(n)} - \frac{R}{\sqrt{n + 0.5}}
                \frac{p^{(n)}}{\|
            p^{(n)} \|_2}\Big\}, 
            \label{eq:SG_iter_2}
        \end{equation}
\end{enumerate}
where $\pi_Q$ is the Euclidean projection on $Q$, and $R$ is an upper bound on the
diameter of the set $Q$: $\| \beta_1 - \beta_2 \|_2 \leq R, \; \forall
\beta_1,\beta_2 \in Q$. Note that the step-size is decreasing in time, and
hence the effect of the regularization, exactly as we wanted.

\begin{remark}
In terms of the variables of our problem, an upper bound $R$ can be found
explicitly, since
\begin{align}
    \allowdisplaybreaks[4]
    \| \theta_1 - \theta_2 \|^2_2 &= \sum_{x=1}^X \sum_{u=1}^U \sum_{k=0}^N \big(\theta_1(x,u,k)
- \theta_2(x,u,k) \big)^2 \notag \\
&= \sum_{x=1}^X \sum_{k=0}^N \Big( \sum_{u=1}^U \big( \theta_1(x,u,k) - \theta_2(x,u,k)
\big)^2 \Big) \notag \\
&\leq \sum_{x=1}^X \sum_{k = 0}^N 2 \notag \\
&= 2 X (N+1),
\end{align}
for all $\theta_1$ and $\theta_2$ fulfilling the simplex constraint (for each
fixed pair of $x$ and $k$). We thus take $R = \sqrt{2X(N+1)}$.
\end{remark}

For explicit expressions of the subgradients, see the calculations in
\aref{app:calc_subgradients}.

\subsection{Summary of algorithm}

The following scheme illustrates the algorithm:
\begin{equation}
    \overset{\circlearrowright}{\min_{\pi} \tikzmark{b} \text{LP}}
    \rightarrow \min_{p,\theta} \text{NLP}
    = \min_p \{ \min_\theta \text{NLP} \} \leadsto \overset{\circlearrowright}{\min_\theta \tikzmark{a}
    \text{RP}},
    \tikz[overlay,remember picture]{\draw[->,square arrow] (a.south) to (b.south);}
    \vspace{0.2cm}
    \label{eq:algorithm}
\end{equation}
where LP is problem \eref{eq:LP}, the regularized \emph{non-linear problem}
(NLP) is problem \eref{eq:full_regularized} and the \emph{relaxed problem} (RP)
is problem \eref{eq:penalized_problem}. The algorithm first performs
$i_\text{ADMM}$ ADMM updates on the LP using equations \eref{eq:ADMM_iter_1},
\eref{eq:ADMM_iter_2} and \eref{eq:ADMM_iter_3}. It then translates the problem
to the regularized NLP, using relations \eref{eq:theta_def} and \eref{eq:pi2p},
and relaxes it to obtain the RP. In this formulation, $i_\text{SG}$ subgradient
steps are taken in the $\theta$ variable using equations \eref{eq:SG_iter_1}
and \eref{eq:SG_iter_2}. This could be interpreted as a sequential
minimization\footnote{See \cite[p.  133]{boyd_convex_2004}.}, however, instead
of performing the subsequent minimization over $p$, we translate back to the
original LP and repeat.

The convergence of the algorithm is guaranteed by the following theorem.
\begin{thm}
    The iterates obtained using the algorithm \eref{eq:algorithm} will converge
    to an optimal policy.
\end{thm}

\emph{Proof (outline):} Instead of providing a formal proof of the theorem, we give the
following heuristic argument. The LP is globally convergent and the effect of the
subgradient steps in the RP is diminishing over time (due to the iteration dependent
step-size). Hence, after a certain number of iterations, the effect of the subgradient
steps will be negligible and the ADMM steps on the LP will converge due to guarantees on
convergence for ADMM (see \cite{boyd_distributed_2010}). \hfill $\square$

It should be noted that after a certain point in time, the subgradient updates will be
pure delays in the algorithm (since the step-size is essentially zero). Hence, it could be
motivated to switch to using plain ADMM after a pre-defined number of iterations and only
use the proposed method as an initial boost. This would reduce the proof to simply
convergence of plain ADMM on an LP.

\section{Examples}
\label{sec:examples}

In this section, we present conditions and several examples of MDPs that have
monotone optimal policies. We also provide numerical simulations illustrating
the performance of the proposed algorithm.

\subsection{Markov decision processes with monotone policies}
\label{sec:monotone}

We start by stating formal conditions which guarantee the existence of
a monotone optimal policy. The following result and four assumptions are
well-known, see, e.g., \cite{krishnamurthy_partially_2016} or
\cite{puterman_markov_1994}:
\begin{enumerate}[leftmargin=1.2cm,label=\textbf{(A\arabic*)}]
    \item Costs $c(x,u,k)$ are decreasing in $x$. The terminal cost $c_N(x)$ is
        decreasing in $x$.
    \item $P_i(u,k) \leq_s P_{i+1}(u,k)$ for each $i$ and $u$. Here $P_i(u,k)$
        is the $i$th row of the transition matrix for action $u$ at time $k$
        and $\leq_s$ denotes first order stochastic dominance, that is,
        $\sum_{i=j}^X P_{i}(u,k) \leq \sum_{i=j}^X P_{i+1}(u,k)$ for all $j \in
        \mathcal{X}$.
    \item $c(x,u,k)$ is submodular in $(x,u)$ at each that $k$. That is,
        $c(x,u+1,k)-c(x,u,k)$ is decreasing in $x$.
    \item $P_{ij}(u,k)$ is tail-sum supermodular in $(i,u)$, i.e., $\sum_{j
        \geq l} (P_{ij}(u+1,k) - P_{ij}(u,k))$ is increasing in $i$.
\end{enumerate}
Note that these four conditions are easily checked. If they are satisfied, then
the following structural result holds:
\begin{thm}
    Assume that an unconstrained finite-horizon MDP satisfies conditions
    \textbf{(A1-4)}. Then there exists a monotone optimal policy.
    \label{thrm:monotone_policy}
\end{thm}

Even though the assumptions \textbf{(A1-4)} might sound restrictive at first
sight, a large class of real-world problems satisfies them. This is because they
are often fulfilled in problems where a degradation takes place over time. To
get some intuition of when the assumptions might hold, we provide the following simple,
but representative, machine replacement example.

Let $\mathcal{X} = \{1, 2\}$ represent the two states of a machine: 1 - broken,
2 - working. Let $\mathcal{U} = \{1, 2\}$ be the two actions: 1 - replace,
2 - continue operation. Let $\theta$ be the probability of a working machine breaking down.
The transition probability matrices are hence:
\begin{equation}
    P(1) = \begin{bmatrix} 0 & 1 \\ 0 & 1 \end{bmatrix}, \quad
P(2) = \begin{bmatrix} 1 & \;\, 0 \\ \theta & \;\, 1 - \theta \end{bmatrix}.
\end{equation}
Let $R \geq 0$ be the cost of performing a replacement (regardless of the state of
the machine) and $\gamma \geq 0$ be the cost of not being able to utilize the
machine because it is broken. Define the costs as 
\begin{equation}
    c(1) = \begin{bmatrix} R \\ R \end{bmatrix}, \quad
    c(2) = \begin{bmatrix} \gamma \\ 0 \end{bmatrix}.
\end{equation}
It is easily checked that this system fulfills conditions \textbf{(A1-4)}. An
optimal policy corresponds to the optimal choices of when to replace the
machine, depending on the current time and its current state, as to maximize
the profits of the operator.

This model can be generalized to larger and more complex systems (e.g., with
time-varying parameters). A recent example of this is medical treatment
planning of abdominal aortic aneurysms, see \cite{mattila_markov_2016}, where
the parameters are time-varying and the optimal policy is monotone.

A number of real-world examples of monotone MDPs (e.g., inventory models,
queueing control, price determination and equipment replacement) can be found
in \cite{puterman_markov_1994}. \cite{krishnamurthy_partially_2016} provides
several examples, including the constrained case, of, e.g., transmission
scheduling over wireless channels. \cite{jiang_approximate_2015} contains an
extensive overview of applications in operations research, energy, healthcare,
finance and economics, that have a monotone structure.

\subsection{Numerical performance}

To illustrate the performance of the proposed method, we generated a synthetic
MDP of dimensions $X = 10$ and $U = 3$ by randomly sampling a system from the
systems that fulfill assumptions \textbf{(A1-4)}. The time-horizon in the MDP
was set to $N = 365$. We will first discuss our rationale for our numerical
choices of the four tuning parameters: $\lambda$, $\rho$, $i_\text{ADMM}$ and
$i_\text{SG}$.

First, the regularization parameter $\lambda$ was chosen as to approximately
balance the regularization term with the current cost. In particular, it was
chosen as the time-horizon times the mean (in time, state and action) of
the cost function, i.e.,
\begin{equation}
    \lambda = \frac{1}{XU} \sum_{x\in \mathcal{X}} \sum_{u\in \mathcal{U}} \Bigg( \sum_{k=0}^{N-1}
    c(x,u,k) + c_N(x) \Bigg).
\end{equation}

We run $i_\text{ADMM} = 10$ ADMM iterations and $i_\text{SG} = 5$ subgradient
steps. Note that we cannot choose a too large value of $i_\text{SG}$ since $p$
is assumed to be constant in the relaxed problem \eref{eq:penalized_problem}.

It is \emph{a priori} difficult to know what the optimal value of $\rho$ is. To
explore the influence of $\rho$ on the problem, we solved the problem for
a range of values between 0.1 and 100 -- see \tref{tbl:performance_rho} and the
discussion below. Note that this is not a feasible approach in a real problem
since one does not want to re-solve the problem. The optimal $\rho$ appears to
be in the lower region of the scale, however, in practice one would perhaps end
up with picking a bigger value.

\begin{figure}[t!]
    \captionsetup[subfigure]{justification=centering}
\begin{subfigure}[]{1.0\columnwidth}
    \centering
    \begin{tikzpicture}[xscale=1.05]
      \begin{axis}[
          title=Cost,
          width=\linewidth, 
          grid=major, 
          grid style={dashed,gray!30}, 
          xlabel=iteration, 
          ylabel=$|c^{(n)} - c^*|$,
          x post scale=0.9, 
          y post scale=0.50, 
          enlarge x limits=false, 
          legend style={
                        },
        ]

        \def\iterADMM{10},
        \def\iterSG{5},
        \def\iterMAX{250},

        \pgfplotsinvokeforeach{0,\iterADMM + \iterSG,...,\iterMAX}{
            \fill [pattern=crosshatch, fill opacity=0.10]
                (axis cs:#1, \pgfkeysvalueof{/pgfplots/ymin}) 
                    rectangle
                (axis cs:{#1+\iterADMM}, \pgfkeysvalueof{/pgfplots/ymax});
        }

        \addplot[blue, thick] 
            table[x=iter,y=ADMM_delta_cost,col sep=comma,
                  select coords between index={1}{175}] {figures/benchmark_data.csv}; 
        \addplot[red, densely dashed, thick]
            table[x=iter,y=ADMM_SG_delta_cost,col sep=comma,
                  select coords between index={1}{175}] {figures/benchmark_data.csv}; 
        \legend{Plain ADMM, Proposed method}
      \end{axis}
    \end{tikzpicture}
    \caption{Difference between the expected cost of using the policy
    at the current iteration compared to the optimal cost.}
    \label{fig:numerical_cost}
\end{subfigure}
\begin{subfigure}[]{1.0\columnwidth}
    \centering
    \begin{tikzpicture}[xscale=1.05]
      \begin{axis}[
          title=Primal residual,
          width=\linewidth, 
          grid=major, 
          grid style={dashed,gray!30}, 
          xlabel=iteration, 
          ylabel=$\|r^{(n)}\|_\infty$,
          x post scale=0.9, 
          y post scale=0.50, 
          enlarge x limits=false, 
          legend style={
                        font=\fontsize{6}{5}\selectfont}, 
        ]

        \def\iterADMM{10},
        \def\iterSG{5},
        \def\iterMAX{250},

        \pgfplotsinvokeforeach{0,\iterADMM + \iterSG,...,\iterMAX}{
            \fill [pattern=crosshatch, fill opacity=0.10]
                (axis cs:#1, -1)
                    rectangle
                (axis cs:{#1+\iterADMM}, \pgfkeysvalueof{/pgfplots/ymax});
        }

        \addplot[blue, thick] table[x=iter,y=ADMM_prim_res,col sep=comma, select coords between index={1}{175}] {figures/benchmark_data.csv}; 
        \addplot[red, densely dashed, thick] table[x=iter,y=ADMM_SG_prim_res,col
        sep=comma, select coords between index={1}{175}] {figures/benchmark_data.csv}; 
      \end{axis}
    \end{tikzpicture}
    \caption{The primal residual (a measure of the feasibility of the policy).}
    \label{fig:numerical_residual}
\end{subfigure}

\caption{Gray regions indicate when ADMM updates are performed, and white
    regions when subgradient steps are taken on the regularized problem (in
    the proposed method).}
    \label{fig:numerical}

\end{figure}
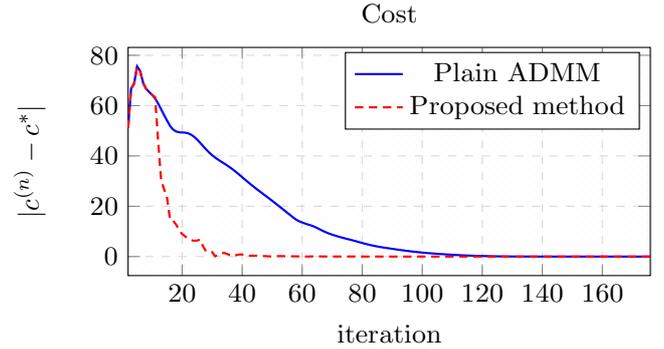
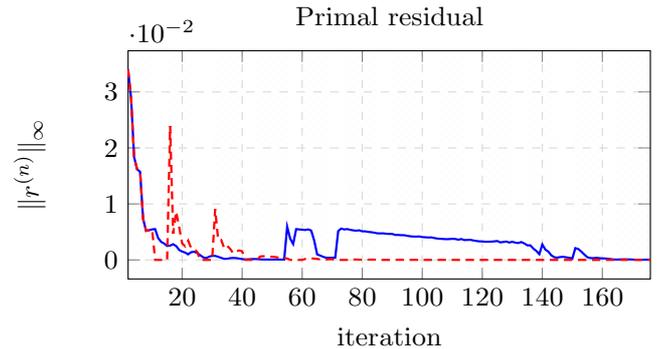

The typical performance of the proposed algorithm (dashed red), as well as plain ADMM
(solid blue), can be seen in \fref{fig:numerical}. A slightly higher value ($\rho
= 30$) than the optimal was chosen for $\rho$. The cost-plot
(\fref{fig:numerical_cost}) shows the
difference in expected cost inquired using the policy at each iteration
compared to using the optimal policy. The residual-plot
(\fref{fig:numerical_residual}) shows the
$l_\infty$-norm of the primal residual which is an indication of how feasible the
policy is in terms of the constraints (e.g., transitions and sum-to-one). Note
that the primal-residual is formulated in terms of the ADMM variables and is
not calculated when the subgradient steps are performed. The areas with gray 
background indicate where ADMM updates are made, and the white areas indicate
where subgradient steps are taken on the regularized problem.

From \fref{fig:numerical}, it is clear that the proposed algorithm steers the iterates
towards the optimum, as seen by the decreases in the cost function when the regularized
problem is used. In early iterations, the iterates become more infeasible when changing
back to ADMM due to the simplifications done to arrive arrive at problem
\eref{eq:penalized_problem} -- for example, assuming $p$ to be constant. At some point in
time, switching between the two formulations can become problematic due to conditioning on
highly unlikely events -- c.f. equation \eref{eq:theta_def}. A work-around is to switch
back to pure ADMM after some fixed number of iterations and use the regularized problem
only as an initial boost -- this is explored in \aref{app:initial_boost}.

\begin{table}[t!]
    \setlength{\captionwidth}{1.0\columnwidth}

    \centering
    \begin{tabular}{c||c|c|c|c}
    ~   & \multicolumn{2}{c|}{Plain ADMM} & \multicolumn{2}{c}{Proposed method} \\
        $\rho$ & \hspace{-2.5mm} $\|r^{(n)}\|_\infty \! < \! \varepsilon_r$ \hspace{-2.5mm}
        & \hspace{-2.5mm} $\frac{|c^{(n)} - c^*|}{c^*}
        \! < \! \varepsilon_c$ \hspace{-2.5mm} & \hspace{-2mm} $\|r^{(n)}\|_\infty
        \! < \! \varepsilon_r$ \hspace{-2.5mm} & \hspace{-2.5mm} $\frac{|c^{(n)} - c^*|}{c^*}
        \! < \! \varepsilon_c$ \hspace{-2mm} \\
    \hline
    0.1 & $>$250 & 191 & $>$250 & $>$250 \\
    1.0 & 94 & 50 & 137 & 76 \\
    5.0 & 68 & 22 & 71 & 31 \\
    10.0 & 82 & 31 & 94 & 21 \\
    20.0 & 118 & 64 & 70 & 27 \\
    30.0 & 169 & 96 & 77 & 28 \\
    40.0 & 212 & 128 & 78 & 31 \\
    50.0 & 246 & 160 & 79 & 31 \\
    60.0 & $>$250 & 192 & 93 & 31 \\
    70.0 & $>$250 & 224 & 101 & 31 \\
    80.0 & $>$250 & $>$250 & 116 & 31 \\
    90.0 & $>$250 & $>$250 & 129 & 31 \\
    100.0 & $>$250 & $>$250 & 143 & 31 \\
    \end{tabular}
    \vspace{0.2cm}
    \caption{Influence of the parameter $\rho$ on the number of iterations
        needed to definitively reach the tolerance in cost and feasibility. The
        tolerance on the relative error in cost was 1\% and the tolerance on the residual was $10^{-4}$.}
    \label{tbl:performance_rho}
\end{table}

It is seen from \fref{fig:numerical} that roughly half the number of iterations are needed
using the proposed algorithm, compared to plain ADMM.  A quantitative comparison is made
in \tref{tbl:performance_rho}. There, both plain ADMM and the proposed algorithm were run
for a fixed number of iterations. The iteration number after which predefined thresholds
held in both terms of cost and feasibility were recorded. In cost, we required the
relative error to be less than 1\%, i.e., $\frac{|c^{(n)} - c^*|}{c*} < \varepsilon_c$,
where $c^{(n)}$ is the expected cost from an initial state using the policy at
iteration $n$, $c^*$ is the expected cost using the optimal policy and $\varepsilon_c$ is
the tolerance of 1\%. In terms of feasibility, we put a threshold on the $l_\infty$-norm
of the residual as $\|r^{(n)}\|_\infty < \varepsilon_r$, for a threshold value of
$\varepsilon_r = 10^{-4}$.

It is apparent from \tref{tbl:performance_rho} that the optimal value of $\rho$
is in the lower region of the scale --- roughly, between 5.0 and 10.0. When
$\rho$ is chosen optimally, the regularization step does not appear to make
much of a difference in terms of convergence speed (for a value of $\rho
= 5.0$, three more iterations are needed to fulfill the criterion). However, if
$\rho$ is chosen suboptimally, the number of iterations needed to fulfill the
convergence criterion is ranging from 59\% to 32\% (and less).

The proposed algorithm converges in roughly the same number of iterations for
relatively high values of $\rho$, as plain ADMM does for the optimal value of
$\rho$: 82 iterations for $\rho = 10.0$ (ADMM) versus 93 iterations for $\rho
= 60.0$ (proposed method). This indicates that the regularization gives
a robustness in the choice of the tuning parameter $\rho$. Since the optimal
value of $\rho$ is unknown to start with, using the regularization allows
for a (larger) suboptimal value to be chosen without much loss in performance.


\section{Conclusions}
\label{sec:conclusions}

This paper has presented a method to accelerate the search for an optimal
monotone policy in MDPs where such a policy is present, exploiting its 
inherent sparsity. A technique from the field of sparse estimation,
namely, \emph{nearly-isotonic regression}, was used as a regularizer to
promote monotonicity in the optimization iterates. To ensure both convergence
and acceleration, two problem formulations were employed: one globally
convergent LP formulated in terms of occupation measures, and a relaxed
regularized problem formulated in terms of conditional probabilities. Numerical
simulations showed the possibility of improvement in terms of number of iterations needed
for convergence when combined with a popular large-scale optimization algorithm
-- especially when a tuning parameter, of \emph{a priori} unknown optimal
value, was chosen suboptimally. 

In the future, it would be of interest to consider memory efficient
representations and how splitting methods could be employed for distributed
computing. It would also be of interest to extend the work to \emph{partially
observed} MDPs.

\bibliography{rob_references.bib}
                                                   
\appendix

\balance

\section{Computation of Subgradients}
\label{app:calc_subgradients}

Here, we compute the subgradients needed in equation \eref{eq:SG_iter_1},
i.e., when updating the iterate on the regularized problem. Compared to the
nominal problem \eref{eq:subgradient_nominal}, we have that:
\begin{itemize}
    \item $\beta = \theta \in \Rb^{X \times U \times (N+1)}$,
    \item $Q = \{ \theta | \sum_u \theta(x,u,k) = 1\}$, i.e., a simplex for
        every pair of $x$ and $k$,
    \item $f(\beta) = f(\theta) = f_1(\theta) + f_2(\theta)$, where
        \begin{align*}
            f_1(\theta) = \sum_{x \in \mathcal{X}} \sum_{u
            \in \mathcal{U}} \Big\{ \sum_{k=0}^{N-1} &c(x,u,k) \theta(x,u,k) p(x,k) \\
    &+ c_N(x) \theta(x,u,N) p(x,N) \Big\},
\end{align*}
and
    \begin{align*}
    f_2(\theta) &= \lambda \sum_{k=0}^N \sum_{x=1}^{X-1} \Big\{
    [1\;2\;\cdots\;U] \big( \theta(x,:,k) \\
    & \hspace{4cm} - \theta(x+1,:,k) \big) \Big\}_+ \\
    &= \lambda \sum_{k=0}^N \sum_{x=1}^{X-1} \Big\{ \sum_{u=1}^U
    \big( \theta(x,u,k) - \theta(x+1,u,k) \big) \Big\}_+.
    \end{align*}
    \item $\bar f(\beta) = \bar f(\theta) = \max_{x,u,k} \{ - \theta(x,u,k)\}.$ 
\end{itemize}
We now need to evaluate a subgradient for each one of these functions.  We have
that $g = g_1 + g_2$ is one subgradient, where
\begin{itemize}
    \item $g_1 = \frac{\partial f_1(\theta)}{\partial \theta_{x',u',k'}}
        = c(x',u',k') p(x',k')$ if $k < N$, and $c_N(x') p(x',k')$
        otherwise.
        \vspace{0.3cm}
    \item \begin{align*}
            g_2 &= \frac{\partial f_2(\theta)}{\partial \theta_{x',u',k'}}
            \notag \\
            &=
            \frac{\partial}{\partial \theta_{x',u',k'}} \lambda \sum_{k=0}^N
            \sum_{x=1}^{X-1} \Big\{ \sum_{u=1}^U \big( \theta(x,u,k) \\
            & \hspace{4cm} - \theta(x+1,u,k) \big) \Big\}_+ \notag \\
            &= \lambda \sum_{k=0}^N
            \sum_{x=1}^{X-1} \ind{\sum_{u=1}^U \big( \theta(x,u,k)
            - \theta(x+1,u,k) \big) > 0 } \\
            & \hspace{1.3cm} \times \frac{\partial}{\partial \theta_{x',u',k'}} \sum_{u=1}^U \big( \theta(x,u,k) - \theta(x+1,u,k) \big) \notag \\
            &= \lambda \sum_{k=0}^N
            \sum_{x=1}^{X-1} \ind{\sum_{u=1}^U \big( \theta(x,u,k)
            - \theta(x+1,u,k) \big) > 0 } \\
            & \hspace{1.3cm} \times \sum_{u=1}^U \big(
            \ind{x=x',u=u',k=k'} \\
            & \hspace{1.3cm} - \ind{x+1=x',u=u',k=k'} \big) \notag \\
            &= \lambda \ind{\sum_{u=1}^U \big( \theta(x',u,k')
            - \theta(x'+1,u,k') \big) > 0 } \\
            & - \lambda \ind{\sum_{u=1}^U \big(
                \theta(x'-1,u,k')
            - \theta(x',u,k') \big) > 0 },
        \end{align*}
        where the first term is only active when $x' < X$ and the last term is
        only included when $x' > 1$.
    \item Let $\bar x, \bar u, \bar k$ be such that
        \begin{equation*}
            \theta(\bar x, \bar u, \bar k) = \argmax_{x,u,k} \{ -\theta(x,u,k)
            \}.
        \end{equation*}
        Then one subgradient is given by
        \begin{equation*}
            \bar g = \frac{\partial \bar f(\theta)}{\partial \theta_{x',u',k'}}
            = -\ind{x'=\bar x, u'=\bar u, k'=\bar k}.
        \end{equation*}
\end{itemize}

\section{Initial Boost}
\label{app:initial_boost}

\begin{figure}[b!]
    \captionsetup[subfigure]{justification=centering}
\begin{subfigure}[]{1.0\columnwidth}
    \centering
    \begin{tikzpicture}[xscale=1.05]
      \begin{axis}[
          title=Cost,
          width=\linewidth, 
          grid=major, 
          grid style={dashed,gray!30}, 
          xlabel=iteration, 
          ylabel=$|c^{(n)} - c^*|$,
          x post scale=0.9, 
          y post scale=0.60, 
          enlarge x limits=false, 
          legend style={
                        font=\fontsize{6}{5}\selectfont}, 
          ymode=log,
          extra x ticks={75},
          extra x tick style={    
                    xticklabel pos=right,   
                    xmajorgrids=true            
                }
        ]

        \def\iterADMM{5},
        \def\iterSG{3},
        \def\iterMAX{75},

        \fill [fill opacity=0.10] (axis cs:0, 1e-16) rectangle (axis cs:75, 250);

        \addplot[blue, thick] table[x=iter,y=ADMM_delta_cost,col sep=comma] {figures/testing_larger/benchmark_data.csv}; 
        \addplot[red, densely dashed, thick] table[x=iter,y=ADMM_SG_delta_cost,col sep=comma] {figures/testing_larger/benchmark_data.csv}; 

        \legend{Plain ADMM, Proposed method}
      \end{axis}
    \end{tikzpicture}
    \caption{Difference between the expected cost of using the policy
    at the current iteration compared to the optimal cost.}
    \label{fig:IB_numerical_cost}
\end{subfigure}
\begin{subfigure}[]{1.0\columnwidth}
    \centering
    \begin{tikzpicture}[xscale=1.05]
      \begin{axis}[
          title=Primal residual,
          width=\linewidth, 
          grid=major, 
          grid style={dashed,gray!30}, 
          xlabel=iteration, 
          ylabel=$\|r^{(n)}\|_\infty$,
          x post scale=0.9, 
          y post scale=0.60, 
          enlarge x limits=false, 
          legend style={
                        font=\fontsize{6}{5}\selectfont}, 
          ymode=log,
          extra x ticks={75},
          extra x tick style={    
                    xticklabel pos=right,   
                    xmajorgrids=true            
                }
        ]

        \def\iterADMM{5},
        \def\iterSG{3},
        \def\iterMAX{75},

        \fill [fill opacity=0.10] (axis cs:0, 1e-16) rectangle (axis cs:75, 25);

        \addplot[blue, thick] table[x=iter,y=ADMM_prim_res,col sep=comma] {figures/testing_larger/benchmark_data.csv}; 
        \addplot[red, densely dashed, thick] table[x=iter,y=ADMM_SG_prim_res,col sep=comma] {figures/testing_larger/benchmark_data.csv}; 
      \end{axis}
    \end{tikzpicture}
    \caption{The primal residual (a measure of the feasibility of the policy).}
    \label{fig:IB_numerical_residual}
\end{subfigure}

\caption{Performance of the proposed method when, after 75 iterations, only plain ADMM
updates are employed.}
    \label{fig:IB_numerical}

\end{figure}
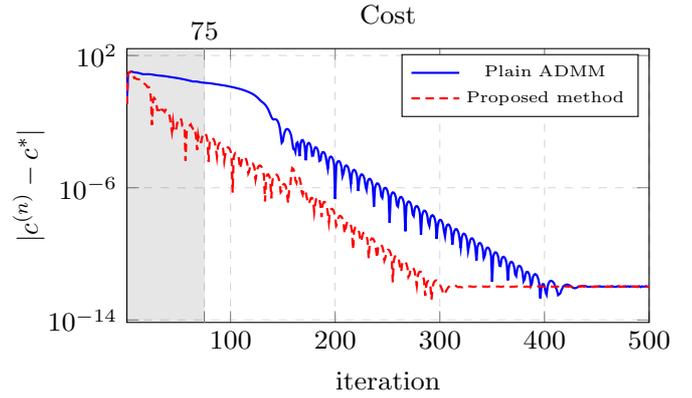
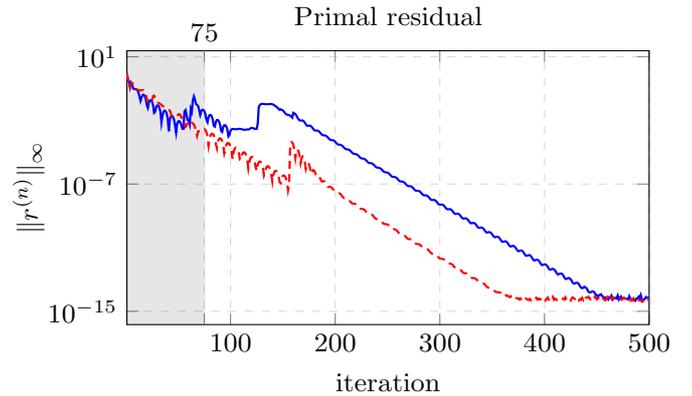

In \fref{fig:IB_numerical}, we consider an example where we use the proposed method only
as an initial boost. In particular, we considered a random (monotone) MDP of size $X = 10$
and $U = 3$ over a time-horizon $N = 60$. The parameters of the method were set to $\rho
= 30$, $i_\text{ADMM} = 5$ and $i_{SG} = 3$.

\end{document}